\documentclass{aac}
\usepackage{graphicx}
\usepackage{alltt}
\usepackage{wasysym}
\usepackage{amssymb}

\usepackage{txfonts}

\begin{document}

\title{Tidal evolution of CoRoT massive planets and brown dwarfs and of their host stars}

\author{S. Ferraz-Mello}      
\institute{Instituto de Astronomia, Geof\'{\i}sica e Ci\^encias Atmosf\'ericas (IAG), Universidade de S\~ao Paulo, Brasil \\              \email{sylvio@iag.usp.br}
   }    

\abstract
  %
{} 
  %
{Revisit and improvement of the main results obtained in the study of the tidal evolution of several massive CoRoT planets and brown dwarfs and of the rotation of their host stars.} 
   %
{Simulations of the past and future evolution of the orbital and rotational elements of the systems under the joint action of the tidal torques and the braking due to the stellar wind.}
   %
{Presentation of several paradigms and significant examples of tidal evolution in extrasolar planetary systems. It is shown that the high quality of the photometric and spectrographic observations of the CoRoT objects allow for a precise study of their past and future evolution and to estimate the tidal parameters ruling the dissipation in the systems.}
  %
{}


\maketitle
%

\section{Introduction}

The CoRoT exoplanets and brown dwarfs are, by far, the most suitable extrasolar planetary systems for the appraisal of the consequences of the exchanges and losses of mechanical energy and angular momentum associated with the tidal deformation of the bodies. 

Given the almost impossibility of the direct observation of the very slow tidal changes in the orbital and rotational elements,
the only way to assess the strength of these changes is by comparing the current values of those elements to their assumed primordial values. For that sake, the main tracers of the past tidal effects in the observed extrasolar planetary systems are  the stellar rotational period and the orbital eccentricity. One element strongly affected by tides is the rotation of the exoplanet; however, besides not observed, the rotation capture into a steady supersynchronous state (a.k.a. pseudo-synchronous) predicted by classical tidal theories, is very fast and close-in exoplanets are expected to be in that state with no memory of its past evolution. 

The privilege of the CoRoT exoplanets over others stem from several reasons:
\begin{itemize}
\item 
The better signal-to-noise ratio of space photometric measurements;
\item
The careful follow-up observations using last-generation spectrographs. They allowed for the determination of the true planetary masses and for the accurate modelling of the host stars, which resulted in the reasonable knowledge of their physical characteristics and ages; 
\item 
The better orbital eccentricities resulting from the joint analysis of transits and radial velocities observed in close epochs, instead of estimations based only on the asymmetry of the radial velocity curves when transits are not observed;
\item
The continuous photometry over long timespans, in many cases allowing for the determination of the stellar rotational periods without the inclination indetermination of the spectroscopically measured $V_{\rm rot}.\sin I$.
\end{itemize} 

When all these elements and parameters are available, the existing theories may be used to have an estimation of the past evolution of the system and the ruling out of scenarios in contradiction with our knowledge of planetary and stellar formation and of the planetary migration, may be used to constrain the tidal parameters entering in the theories. 
This is a fundamental contribution because the currently used tidal parameters of exoplanets and their host stars are just extrapolations from solar system analogs and the imprecision in their knowledge do not allow us to make reliable predictions of the future evolution of these systems. The actual knowledge of the real parameters is currently limited to the results issued from simulations of the distribution of the initial eccentricity values of close-in planets (Jackson et al. 2008), and from the analysis of the survival of some short-period planets around evolving stars and of the limits for circularization of planets with highly eccentric orbits (Hansen, 2012).

One extra factor favoring the use of the CoRoT systems in the study of tidal phenomena is the great proportion of very massive planets and brown dwarfs among them. About 1/3 of the discovered CoRoT planets have masses larger than 2 M$_{\rm Jup}$. The large masses of the companions enhance tidal effects and thus allow for more robust analyses. 

\begin{table*}


\caption{Basic parameters of the CoRoT systems discussed in this paper}\label{table}

\begin{tabular}{lccccccc}  \hline\hline

{Star} & {Spect.} & {Star mass } & {Compan. mass} & {Orbital period} & {Rotation period}  & Eccentricity &
 {Age}\\

         & {type} & {(M$_\odot$)} & {(M$_{\rm Jup})$} & {(d)} & { (d)} && {(Gyr)} \\
\hline

CoRoT-2$^1$ & G7V & $ 0.97 \pm 0.06$&3.31 $\pm$ 0.16 &  1.742996(2) & 4.52 $\pm$ 0.02$^a$  & $<$ 0.06 &  0.2 $-$ 4.0\\
CoRoT-3$^{2,13}$ & F3 & $ 1.37 \pm 0.09$& 21.8 $\pm$ 1.0 &  4.256799(4) & $\sim$ 3.65 $^{a\ [15]}$  & 0.012(10) &  1.6 $-$ 2.8 \\
CoRoT-5$^3$ & F9V & $ 1.0 \pm 0.02$& 0.467 $^{+0.047}_{-0.024}$ &  4.037896(2) & $>$ 30 $^b$  & $0.09^{+0.09}_{-0.04}$ &  6.9 $\pm$ 1.4  \vspace*{0.5mm}\\
CoRoT-11$^{4}$& F6V & $1.27\pm 0.05$  & $2.33 \pm 0.27$  &  2.994325(21) 
& $1.73 \pm 0.22^b$  & $<$ 0.2 &  2.0 $\pm$ 1.0\\
CoRoT-14$^{5,14}$& F9V & $1.13\pm 0.09$  &6.94 $\pm$ 0.5 &  1.51214  & $5.6 \pm 0.5^{a}$  & 0 &  0.4 $-$ 8.0\\
CoRoT-15$^6$& F7V  &1.32$ \pm$ 0.12 & 63.3 $\pm$ 4.1 &  3.06036(3) &   $^c$ & 0&  1.14$-$3.35 \\
CoRoT-16$^7$& G5V  &1.10$ \pm$ 0.08 & 0.535 $\pm$ 0.085 &  5.35227(2) &   $<$ 60  $^b$  &  0.33 $\pm$ 0.1 &  6.7$\pm$2.8 \\
CoRoT-18$^{8,13}$& G9V & $0.95 \pm 0.15$ & 3.47 $\pm$ 0.38 &  1.900070(3) & 5.4 $\pm$ 0.6 $^a$  & $<$0.08  &  0.05-1.0\\
CoRoT-20$^{9}$ & G2V & $ 1.14 \pm 0.08$& 4.24 $\pm$ 0.23 &  2.48553345(7) & 11.5 $\pm$ 3.1 $^b$  & 0.562(13) &  $<$ 1\\
CoRoT-23$^{10}$ &   G0V & $1.14 \pm 0.08 $&  2.8 $\pm$ 0.25&  3.6314(1) & 9.0 $\pm$ 1.0 $^{b, d}$     
& 0.16$\pm$ 0.02  &  $7.2_{-1.0}^{+0.5}$ \\
CoRoT-27$^{11}$ &   G2 & $1.05 \pm 0.11 $&  10.39 $\pm$ 0.55&  3.57532(6) & 12.7 $\pm1.7 ^b$     & $<$ 0.065  &  4.2 $\pm$ 
2.7 \\
CoRoT-33$^{12}$&G9V& $0.86\pm 0.04$ &  $59.0\pm 1.8$ & 5.819143(18)  & $8.936\pm 0.015 ^a $  & 0.0700(16) & $>$ 4.6 \\
\hline

\multicolumn{8}{l}{\textbf {Notes.} (a) photometric; (b) $\times \sin I$; (c) photometric modulations at 2.9, 3.1 and 6.3 d; (d) id. at 10.5 d}\\
\hline               
   
\multicolumn{8}{l}{\noindent\small{ (1) Alonso et al. (2008); (2) Deleuil et al. (2008); (3) Rauer et al. (2009); (4) Gandolfi et al. (2010); (5) Tingley et al. (2011); (6) Bouchy  }} \\
\multicolumn{8}{l}{\small { et al. (2011); (7) Ollivier et al. (2012); (8) H\' ebrard et al. (2011); (9) Deleuil et al. (2012); 
(10) Rouan et al. (2012); (11) Parviainen et al.}}\\
\multicolumn{8}{l}{\small { (2014); (12) Csizmadia et al. (2015); (13) Moutou et al. (2013);  (14) Ferraz-Mello et al. (2015); (15) Tadeu dos Santos (pers.comm.). }}\\
\hline                \end{tabular}             
     
\end{table*}


\begin{figure}
   
\centering
   
\includegraphics[width=7cm]{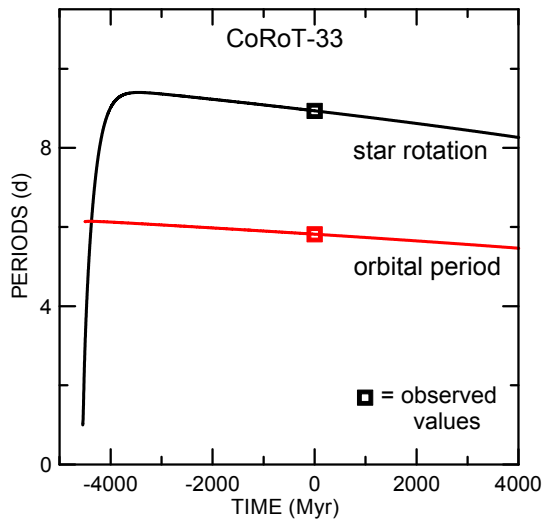}
      
\caption {Simulations of the evolution of the rotational period of CoRoT-33 (black) and of the orbital period of the companion CoRoT-33b (red). Adopted stellar tidal relaxation factor: $\sim$ 36 s$^{-1}$. }
\label{fig:C33}
   
\end{figure}

In this paper, we select among the 30 discovered CoRoT systems those that may be considered as paradigms, or, at least, as very significant examples of tidal evolution scenarios of extrasolar planetary systems. 

In order to model the evolution of the selected systems, we need to use one of the available tidal theories. We adopt the creep tide theory (Ferraz-Mello 2013), for which we dispose of codes allowing us to obtain the variation of the stellar rotational period and of all orbital elements with high formal precision, even when the orbital eccentricities become very large. It is important to stress that all available theories are more or less equivalent when the considered bodies are predominantly fluid, with low Reynolds numbers (Stokes fluids) and should give comparable results to those presented in this paper.
It is also necessary to use one theory to assess the stellar loss of rotational angular momentum through the magnetic phenomena associated with the stellar winds. We adopt the formulas proposed by Bouvier et al (1997). These formulas account in a satisfactory way for the rotation of single stars observed in clusters of known ages (Bouvier, 2013), once the very active initial phases are over, and allows one to reproduce the Skumanich's law (Skumanich, 1972). More complete theories exist taking into account more complex patterns of the magnetic activity of the stars, but they involve physical parameters not known for the considered stars (see Matt et al. 2012).

The  theories used here are the same that were used to study the interplay of tidal evolution and stellar wind braking in the rotation  of stars hosting massive close-in planets by Ferraz-Mello et al. (2015)

\section{CoRoT 33. The paradigm}

All favorable circumstances are present in this system formed by an active G star and a massive brown dwarf in a close-in orbit. The modelling of the tidal evolution of this system unravelled a typical scenario which, in less marked circumstances, happens also in systems with massive planets as CoRoT-27 and KEPLER-75 (see Ferraz-Mello et al., 2015). In its young ages, once the star is decoupled from the disk and is no longer subject to significant changes, we should expect to find it as a fast rotator. 

The evolution of the stellar rotation is controlled by two effects: the torques due to the tides raised by the companion on the star and the leakage of angular momentum due to the magnetic wind braking of the star. Since the magnetic braking $\dot\Omega$ is proportional to $\Omega^3$, the braking dominates the initial evolution of the stellar rotation, when the angular rotational velocity $\Omega$ is large. However, as the age and the rotational period grow, the braking weakens and, after some time, the tidal effects dominate over the braking and the stellar rotation is accelerated. The energy spent accelerating the star rotation leads the planet to spiral towards the star and, if it lives long enough, it will fall on the star, transferring to it all of its angular momentum.
In the case of CoRoT-33, the age of the star is large and it allowed the system to reach this latter stage. The brown dwarf is slowly falling towards the star, the rotation of which is being slowly accelerated. 

Fig. \ref{fig:C33} shows the expected evolution of CoRoT-33 during its lifetime. The estimated stellar tidal relaxation factor $\gamma_{\rm st}$ for this star, depending on the actual orbital and rotational parameters used, lies between 30 and 60 s$^{-1}$. The  formulas relating the tidal relaxation factor and the quality factor $Q$ (see Ferraz-Mello, 2013) indicate that this result corresponds to $Q$ in the interval $3-6\times 10^6$, which is in the mid of the interval 
$7\times 10^5 - 4.5 \times 10^7$ that we obtain from Hansen's (2012) results for a sample of planet hosting stars, if we adopt the stellar Love number $k_2=0.2$.

Several experiments were done letting the companion's rotation to vary freely. The results showed that in that case, the companion takes about 1-2 Gyr to reach a synchronous rotation. However the results for the evolution of the stellar rotation and of the orbital elements ($a,e$) of the companion are the same as those obtained when the capture in the steady supersynchronous stationary rotation is assumed ab initio.

We may mention that the past evolution of the system is critically dependent on the current rotational period. If $\gamma_{\rm st}$ is different, we may have either a star whose rotation was recently too fast or too slow. However, the numerical accuracy required to determine $\gamma_{\rm st}$ is not real. The spectroscopically determined rotational period has large dispersion ($\sim 1$ d) and the photometric period may be different from the actual rotational period. Some striking commensurabilities between the stellar rotational period and the orbital period in some extrasolar planetary systems (e.g. 3/2 in the present case), allow us to suspect that the magnetic fields controlling,  on the surface of the stars, the features responsible for the measured periodic  variations in the star's light are modulated by the orbital period of the massive close-in companions (see B\'eky, 2014).

To conclude, let us add that the comparison of the solutions with and without the consideration of the magnetic braking shows that the circularization of the orbit of the companion and its fall toward the star are both accelerated by the angular momentum leakage since, in this case, the stellar energy losses are also contributing to the planet fall.

\subsubsection*{CoRoT-27}
CoRoT-27 is very similar to CoRoT-33. Aged and with a super-Jupiter as companion, the same evolutionary processes seen in CoRoT-33 are present in this case (see Ferraz-Mello et al. 2015). 
The best representation of the parameters of this system was obtained assuming $\gamma_{\rm st} = 90$ s$^{-1}$ (corresponding to $Q=2.7\times 10^6$ if $k_2=0.2$).  

\subsubsection*{CoRoT-23}
Another system showing a similar behavior is CoRoT-23. The study of the evolution of the stellar rotation with the data given in Table 1 results in $\gamma_{\rm st} \sim 8\ {\rm s}^{-1}$; This value corresponds to the quality factor $Q=4\times 10^5$ (if $k_2=0.2$), which is below the lower limit of the interval determined by Hansen (2012). The discovery paper (Rouan et al. 2012) mentions a peak in the power spectrum of the photometric measurements at 10.5 d. If this value is adopted for the rotational period, the resulting $\gamma_{\rm st}$ and $Q$ are doubled and are in agreement with Hansen's determinations.

\begin{figure}
   
\centering
   
\includegraphics[width=7cm]{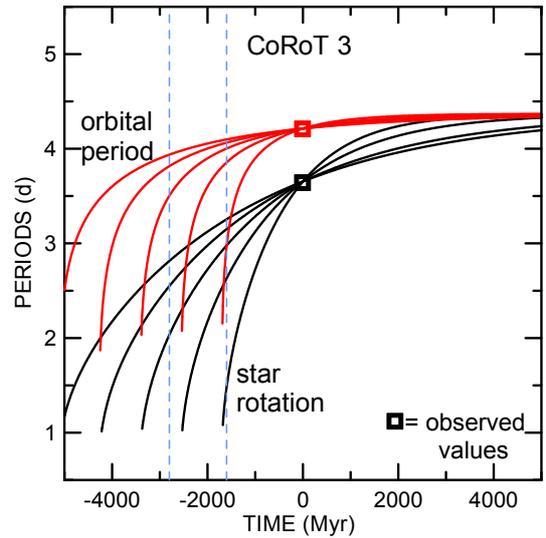}
      
\caption {Simulations of the evolution of the rotational period of CoRoT-3 (black) and of the orbital period of the companion CoRoT-3b (red). Adopted stellar tidal relaxation factors (starting from the steepest curves): 40, 60, 80, 100, 120 s$^{-1}$. The dashed vertical lines show the stellar age range.}
\label{fig:C3}
   
\end{figure}

\section{CoRot 3. The dissipation in an F star}

CoRoT-3b is a brown dwarf in orbit around an F3 star. In all other systems involving a nonactive F star with a massive companion (e.g. CoRoT-15, KELT-1, HAT-P-2), the stellar rotational period is very close to the orbital period of the companion, what means that the system has lost completely the memory of its past evolution and do not allow even for a rough estimate of the stellar tidal dissipation factor $\gamma_{\rm st}$. 
In the case of CoRoT-3, however, the stellar rotation is faster than the orbital period, indicating that the synchronization has not yet been reached. The value 3.65 d obtained by Tadeu dos Santos (comm. pers.) from a reanalysis of the Rossiter-McLaughlin effect in the observations done by Triaud et al. (2009), is reinforced by the existence of a peak at about 3.6 days in the power spectrum of the photometric signal. 

We present on Fig. \ref{fig:C3}, the tracks corresponding to stellar relaxation factors between 40 and 120 s$^{-1}$. In all cases, the star has a fast initial rotation and the brown dwarf is in an orbit above the synchronous (or corotation) orbit. The strong tidal interaction between the companion and the star slows down the star's rotation, which tends to approximate the synchronization. The companion, on its turn, is being accelerated by the tidal torque due to the fast rotating star and earns energy, moving away from the star. 

We may take advantage of the large mass of the brown dwarf and select the evolutionary tracks compatible with the age of the system and its rotational period. The curves leading to a fast stellar rotation in a time corresponding to the age of the system (1.6 -- 2.8 Gyr) are those for which $\gamma_{\rm st}$ is between 40 and 70 s$^{-1}$, which corresponds to $Q \sim 6-9 \times 10^6 $ (if $k_2$=0.2). Given that we have used a planar model and that the study of the Rossiter-McLaughlin effect indicates an obliquity larger than 30 degrees, it is prudent to bracket this value with a somewhat larger interval. The value is of the same order as the values found for G stars (as CoRoT-33 and CoRoT-27, for example), and their comparison serves to estimate the role played in the stellar tidal dissipation by the existence of a convective zone. 

The system may be evolving towards a double synchronization as already indicated by previous studies of this star by Carone (2012). The dissipation in the planet is negligible; tests with several values of the planetary tidal relaxation factor were done and did not show any change in the results.

\subsubsection*{CoRoT-11}

The rotational evolution of the F6V star CoRoT-11 has been studied by Lanza et al. (2011) with a model adopting a magnetic braking of about 1/4 of the value adopted by Bouvier et al. (1997) for less massive stars.
That study allowed them to constrain the value of the stellar tidal quality factor averaged over the life of the system to the interval 
$5\times 10^5 - 2.5 \times 10^6$ (for $k_2=0.2$). 
These values match the lower part of the interval determined by Hansen (2012) for general host stars.
Lanza et al. (2011) also note that their results are $3-15$ times smaller than the prediction from the theoretical results of Barker and Ogilvie (2009) for the quality factor of a similar star.

The study of this system is made particularly difficult by the uncertainty on the current eccentricity of the system.  Lanza et al (2011) adopted $e=0$ on the grounds that a circular orbit is compatible with the available radial velocity measurements.

\section{CoRoT 15. The double synchronization}

Because of its mass and spectral type (F7V), CoRoT-15 is not expected to be active and the evolution may be fully controlled by the tidal interactions. In the past, this system may have shown a situation similar to that of CoRoT-3 (see Fig. \ref{fig:C3}), but in the case of CoRoT-15, because of the larger mass of the companion and its closer proximity to the star, 
this process was faster and the full synchronization may have been reached a long time ago. 
It may be currently in a state in which the eccentricity is zero and all periods, rotational and orbital, are equal.   
Any attempt to model the recent evolution of the system by assuming that the stellar rotational period is different from the orbital period results into physically unacceptable solutions where the star is showing a too fast or a too slow rotation a short time ago. In addition, no residual eccentricity may have survived the fast damping due to those conditions. 

The only alternative to this scenario is the presence of a residual activity (see Barker and Ogilvie, 2009). 
If a small activity, say, around 10 percent of that of a G star is accepted, the situation is almost the same except that the system will no longer be stationary. The brown dwarf will be slowly falling on the star whose rotation will be accelerated preserving the synchronism. But, even in this case,  the system would not be expected to show significant changes during the star's remaining lifetime.

\section{CoRoT-5. Planetary tide and dissipation}

\begin{figure}
   
\centering
   
\includegraphics[width=7cm]{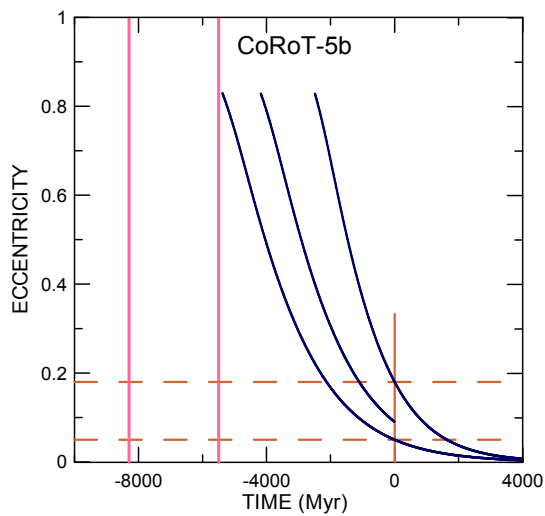}
      
\caption {Simulations of the past tidal evolution of the eccentricity of CoRoT-5b for initial eccentricities 0.05, 0.09 and 0.18. Adopted planetary tidal relaxation factor: 200 s$^{-1}$ The vertical lines show t=0 and the stellar age range. The horizontal lines show  the error bar of the observed eccentricity. (Adapted from Ferraz-Mello, 2013)}
\label{fig:C5}
   
\end{figure}

In this system, the memory of the past evolution is kept by the eccentricity and  it was the first for which the analysis of the discussed models was used to estimate the tidal relaxation factor of one hot Jupiter (Ferraz-Mello, 2013). 
The simulations of this system show a significant rate of circularization and the eccentricity must have been larger in the past. However, if the tidal dissipation is larger than a certain limit, we are forced to assume that the eccentricity has been too high in a recent past. 
All general mechanisms responsible for important eccentricity enhancement are related to events expected to happen only in the early life of the system (see Malmberg \& Davies, 2009). 
Therefore the rate of circularization of the system may not impose high eccentricities in a recent time. Figure \ref{fig:C5} shows the solutions corresponding to the planetary tidal relaxation factor $\gamma_{\rm pl} = 200\ {\rm s}^{-1}$ for initial eccentricities within the range resulting from the analysis of the observations (we used the more probable value and the minimum and maximum of the interval of confidence).
 
The limit $200\ {\rm s}^{-1}$ corresponds to the quality factor $Q=5 \times 10^{6}$ (if $k_2=0.35$)and 
 is inside the range $2\times 10^6 - 2 \times 10^7$ determined by Hansen (2012) from his analysis of survival of close-in exoplanets. However, it is at least one order of magnitude larger than the values that we obtain by extrapolating to the bloated radius of CoRoT-5b, the values of $Q$ determined for the giant planets of our Solar System (see Ferraz-Mello, 2013). It is worth noting that if a relaxation factor 10 times smaller is adopted, the resulting larger circularization rate would impose eccentricities close to 1 in the past Gyr, what would be impossible to explain with the current knowledge of the exoplanets orbital evolution.

The present planetary rotation does not interfere with the results. As extensively studied by Carone (2012), the rotation of the close-in planets in the CoRoT systems is quickly (less than 1 Myr) driven to a stationary solution. In this case, because of the past high eccentricity, the rotation may have been soon captured in a supersynchronous stationary solution and must be evolving since then in the stationary solution often called pseudo-synchronous solution. The stellar rotation, on its turn, is very slow and does not affect the results.

\begin{figure}
   
\centering
   
\includegraphics[width=7cm]{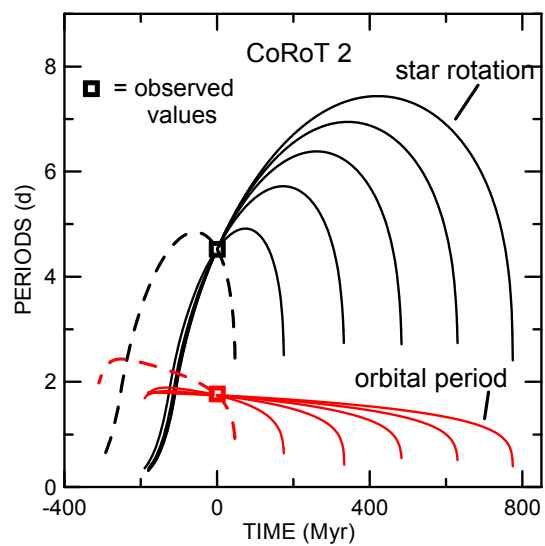}
      
\caption {Simulations of the evolution of the rotational period of of CoRoT-2 (black) and of the orbital period of the planet CoRoT-2b (red). 
Adopted stellar tidal relaxation factors: 5 (dashed line), 20, 40, 60, 80, 100 (solid lines) s$^{-1}$. (Taken from Ferraz-Mello et al., 2015)}
\label{fig:C2}
   
\end{figure}

\section{CoRoT-2. A gyrochronology example}

The host star of this system is likely a very young star. Its X-ray luminosity and strong chromospheric emission indicate a stellar age of about 200-300 Myr (Poppenhaeger \& Wolk, 2014). The low age is the source of the main difficulty. The stellar rotation did not yet reach the point where tidal torques and angular momentum leakage are close to equilibrate themselves, and it is not possible to make an estimation of the stellar tidal relaxation factor. For this reason, we present on Fig. \ref{fig:C2}, for comparison, the tracks corresponding to stellar tidal relaxation factors between 20 and 100 s$^{-1}$. The first noteworthy feature in this fountain-like plot is that the evolution of the past stellar rotation is strongly controlled by the magnetic braking. The role of the tidal forces is secondary and we should assume very strong dissipation (i.e., a very small $\gamma_{\rm st}$) to see a solution not coincident with the dominant ones. The convergence of the paths to the path corresponding to braking only (i.e. to the Skumanich's law) points to the origin of the system 200 Myr ago, exactly as it would result from the application of the  gyrochronology rules without taking the tides into account (see Brown, 2014). The low age of the star brings with it another difficulty. The star is bloated ($R=1.465\pm 0.029 {\rm R}_\odot$) and still contracting. This introduces a new factor in the problem: The rotation has an unmodelled component of acceleration not considered in the used models, and the actual age may be larger than that predicted by gyrochronology rules.

\subsubsection*{CoRoT-18}

Several other stars in the CoRoT family show similar behavior. The most similar case is CoRoT-18 (see Ferraz-Mello et al., 2015) which also is a bloated young star just a bit older than CoRoT-2. 

\subsubsection*{CoRoT-14}

A second case showing some resemblance, but not as young as the previous ones,  is CoRoT-14. In this case, in order to obtain an evolution leading to the current orbital and rotational elements, we would need to assume a much smaller tidal dissipation (much larger $\gamma_{\rm st}$). Several hypotheses were discussed considering this F9V star (see Ferraz-Mello et al. 2015). Besides the possibility of having some actual parameter different of the published ones, we may also suspect that the measured surface rotation does not reflect the rotation of the inner radiative core, which would be rotating much faster.

\begin{figure}
   
\centering
   
\includegraphics[width=7cm]{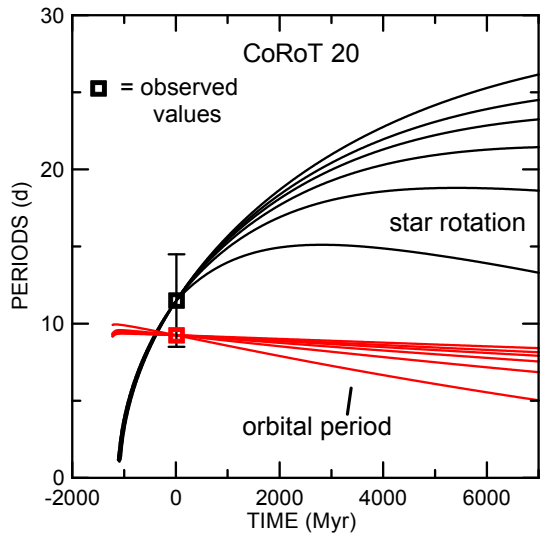}
      
\caption {Simulations of the evolution of the rotational period of of CoRoT-20 (black) and of the orbital period of the planet CoRoT-20b (red). Adopted stellar tidal relaxation factors (starting from the lower curves): 10, 20, 30, 40, 50, 70 s$^{-1}$.}
\label{fig:C20}
   
\end{figure}

\section{CoRoT-20. A high eccentricity system} 

This is also a star of low age in which the angular momentum leakage is dominating. As for CoRoT-2 and CoRoT-18, the age of the star is not large enough to allow us to estimate the value of the stellar tidal relaxation factor $\gamma_{\rm st}$ but, in compensation, the past rotation of the star was dominated by the magnetic brake and the system is well suited for the estimation of its age via gyrochronology. 
As in the case of CoRoT-2, we use a grid of values and construct the fountain-plot shown on Fig. \ref{fig:C20} and see that the evolutionary tracks converge to a fast-rotating star in $\sim 1.1$ Gyr.

The lack of knowledge of the stellar tidal relaxation factor does not allow us any definite prediction of its future evolution.
However, for the expected values of the relaxation factor, the system is stable. The orbit is spiralling towards the star, but at a slow pace. The great difference between this cases and the previous ones is that the current orbit of the companion is very eccentric; the rate of circularization may be significant, but in no case the eccentricity is expected to be close to zero in the star's lifetime.

The rotation of the star will evolve following the patterns described in previous sections. 
It is worth recalling that the previous studies of this system (Deleuil et al. 2011, Carone, 2012) have already pointed out the stability of the system and that,  in the absence of magnetic braking, the system would have been a candidate for future double synchronization

\section{CoRoT-16. A limit case}

\begin{figure}
   
\centering
   
\includegraphics[width=7cm]{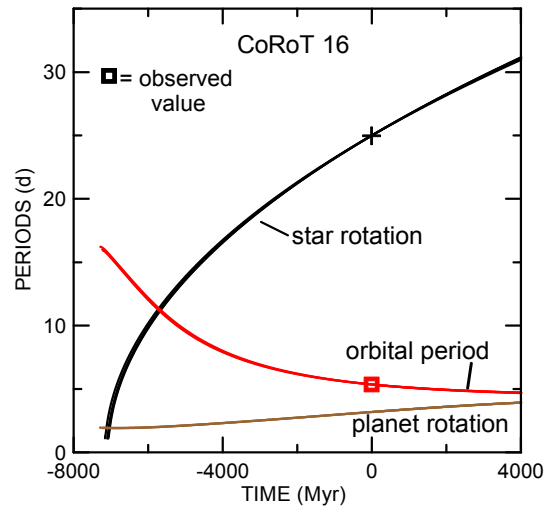}
      
\caption {CoRoT-16: Simulations of the evolution of the stellar (black) and planetary (brown) rotational periods and of the orbital period of the planet (red). Adopted stellar tidal relaxation factors between 30 and 80 s$^{-1}$ (superposed).}
\label{fig:C16}
   
\end{figure}

CoRoT-16b is a hot Saturn in an orbit beyond the limit where the tidal interactions with the star influence significantly the stellar rotation. The large indetermination in the given stellar rotational period does not allow us to use a gyrochronology model to determine the star's age, but if we consider the inverse problem, we see that the known isochronal age of the star may be used to better constrain the current stellar rotational period to the neighborhood of 25 d. 

Fig. \ref{fig:C16} shows the evolution of the three periods present in this system: the orbital period and the rotational periods of both the star and the planet. The thick lines are the superposition of solutions obtained using stellar relaxation factors $\gamma_{\rm st}$ between 30 and 80 s$^{-1}$. The rate of circularization of the orbit is not large and it will last yet some time before circularization, which may happen by the end of the star's lifetime. 

The significant eccentricity of the orbit can be used to constrain the planetary tidal relaxation factor. 
As for CoRoT-5b, the past evolution of the eccentricity depends on the adopted eccentricity. Taking into account the error bar of the current eccentricity, we get, for the limit value, the estimate: $\gamma_{\rm pl} > 30 {\rm s}^{-1}$, which corresponds in terms of the quality factor to  $Q > 9\times 10^5$. This limit is almost equal to the value $8 \times 10^5$ adopted in the discovery paper (Ollivier et al. 2012). It is lower than the limit found for CoRoT-5b. However, this estimate is strongly affected by the error bars in the stellar period and in the orbital eccentricity.
If a more accurate redetermination would be possible, we should expect an improved larger limit.

\section{Conclusion}

This paper shows that several of the CoRoT planetary systems had their tidal evolution studied with some accuracy due to the high-quality of the determined physical parameters, rotational and orbital elements. For these systems, it has been possible to constrain the values of the tidal relaxation factor of the host star or of the planet and the results, with only a partial exception, agree with the estimates of the tidal dissipation quality factor in the extended sample of short-period planetary systems studied by Hansen (2012).

These successful cases may be seen as paradigms or, at least, as significant examples of what can yet be done using other exoplanetary systems. The main perspective concerns systems discovered by KEPLER. Many of these systems are now being observed from ground with last-generation spectrographs and we may expect that these follow-up observations will produce more accurate masses, ages, rotational and orbital elements so that they may be included in the set of successful case studies, as it was already done for KEPLER-75 (see Ferraz-Mello et al. 2015). We may also have these paradigms in mind when planning new missions, as PLATO. With relatively brighter stars as targets, we may hope that the exoplanetary systems to be discovered by PLATO be observed from the ground in epochs not too distant from the epochs of the space photometric observations so that the determination of the eccentricities can actually benefit of the complementarity of the two observational techniques and be obtained with better accuracy.

\begin{acknowledgements}
 
This investigation was supported by grants CNPq 306146/2010-0, FAPESP 2014/13407-4  and by INCT Inespa\c co procs. FAPESP 2008/57866-1 and CNPq 574004/2008-4. 
The help of Mr. E.S. Pereira in the calculation of some complementary examples is acknowledged.

\end{acknowledgements}


\begin{thebibliography}{}

\bibitem[Alonso et al.(2008)]{alo08}
Alonso, R., Auvergne, M., Baglin, A. et al. 2008, 
A\&A,  \textbf{482}, L21 

\bibitem[Barker and Ogilvie (2009)]{Bark}
{Barker, A. J., and Ogilvie, G. I. 2009. }
{MNRAS \textbf{395}, 2268 }   

\bibitem[Beky et al. (2014)]{Bek}
B\'eky, B., Holman, M. J., Kipping, D. M. \& Noyes, R. W.  2014,
ApJ, \textbf{788} id. 1 

\bibitem[Bouchy et al. (2011)]{C15}
Bouchy, F., Deleuil, M., Guillot, T. et al. 2011, 
A\&A, \textbf{525}, A68 

\bibitem[Bouvier et al.(2007)]{Bou}
Bouvier, J., Forestini, M. \& Allain, S.  1997
A\&A, \textbf{326}, 1023     

\bibitem[Bouvier(2013)]{B13}
Bouvier, J.  2013,
EAS Publ. Series Vol. \textbf{62}, 143      

\bibitem[Brown(2014)]{Brw}
Brown, D. J. A.  2014,
MNRAS \textbf{442}, 1844       

\bibitem[Carone, 2012]{Car}
Carone, L. 2012,
Doctoral dissertation, Universit\"at zu K\"oln.

\bibitem[Csizmadia et al. (2015)]{Csiz}
Csizmadia, S., Hatzes, A., Gandolfi, G. et al. 2015
A\&A (in press). Preprint arXiv:1508.05763.

\bibitem[Deleuil et al. (2008)]{Del8}
Deleuil, M., Deeg, H. J., Alonso, R. et al. 2008
A\&A \textbf{491}, 889   

\bibitem[Deleuil et al. (2012)]{Del12}
Deleuil, M., Bonomo, A. S., Ferraz-Mello, S. et al. 2012
A\&A, \textbf{538} A145


\bibitem[Ferraz-Mello et al. (2008)]{FRH}
Ferraz-Mello, S., Rodr\'{\i}guez, A. \& Hussmann, H.  2008,
Celest. Mech. Dyn. Astr.  \textbf{101}, 171 
and Errata: Celest. Mech. Dyn. Astr.: \textbf{104}, 319-320 (2009). (arXiv: 0712.1156)

\bibitem[Ferraz-Mello(2013)]{Rheo}
Ferraz-Mello, S.  2013,
Celest. Mech. Dyn. Astr. \textbf{116}, 109-140 (arXiv: 1204.3957)

\bibitem[Ferraz-Mello et al.(2015)]{hosts}
Ferraz-Mello, S., Santos, M., Folonier, H., Csizmadia, S., do Nascimento Jr, J. D., \&
P\"atzold, M. 2015 
\textit{ApJ} \textbf{807}, id. 78 

\bibitem[Gandolfi et al. (2010)]{Gan}
Gandolfi, D., H\'ebrard, G., Alonso, R. et al. 2010
A\&A \textbf{524} A55

\bibitem[Hansen(2012)]{Han2}
Hansen, B.M.S.  2012,
ApJ \textbf{757}: 6

\bibitem[Hebrard et al.(2011)]{C18}
H\'ebrard, G., Evans, T.M., Alonso, R. et al 2011, 
A\&A \textbf{533} A130

\bibitem[Jackson et al. (2008)]{Jac}
Jackson, B., Greenberg, R. \& Barnes, R. 2008,
ApJ \textbf{678}, 1396

\bibitem[Lanza et al. (2011)]{Lan}
Lanza, A. F., Damiani, C. \& Gandolfi, D. 2011,
A\&A \textbf{529}, A50

\bibitem[MB, 2009]{Malm}
Malmberg, D., Davies, M.B. 2009,
MNRAS \textbf{394}, L26      

\bibitem[Matt et al, 2012]{Matt}
Matt, S. P., MacGregor, K. B., Pinsonneault, M. H., \& Greene, T. P. 2012,
ApJ Letters, \textbf{754}, L26.

\bibitem[Moutou et al.(2013)]{Mut}
Moutou, C., Deleuil, M., Guillot, T. et al. 2013, 
Icarus, \textbf{226}, 1625	

\bibitem[Ollivier et al. (2012)]{Oll}
Ollivier, M., Gillon, M., Santerne, A. et al. 2012
A\&A \textbf{541}, A149.

\bibitem[Parviainen et al. (2014)]{C27}
Parviainen, H., Gandolfi, D., Deleuil, M. et al. 2014, 
A\&A \textbf{562}  A140.

\bibitem[Poppenhaeger and Wolk (2014)]{Pop}
Poppenhaeger, K. \& Wolk, S. J.  2014, 
A\&A, \textbf{565}  L1

\bibitem[Rauer et al. (2009)]{Rau}
Rauer, H., Queloz, D., Csizmadia, S. et al. 2009,
A\&A \textbf{506}, 281      

\bibitem[Rouan et al. (2009)]{Rou}
Rouan, D., Parviainen, H., Moutou, C. et al. 2012,
A\&A, \textbf{537},  A54.

\bibitem[Skumanich (1972)]{Sku}
Skumanich, A.  1972,
ApJ, {\textbf 171}, 565
 
\bibitem[Tingley et al.(2011)]{C14}
Tingley, B., Endl, M., Gazzano, J.-C. et al. 2011,  
A\&A, \textbf{528}, A97

\bibitem[Triaud et al.(2009)]{Tri}
Triaud, A. H., Queloz, D., Bouchy, F. et al. 2009
A\&A  \textbf{506}, 377    
 
\end{thebibliography}
\end{document}